\documentclass[a4paper,12pt]{book}

\usepackage[T1]{fontenc}
\usepackage[utf8]{inputenc}
\usepackage[french,german,english]{babel}

\usepackage[a4paper, twoside, inner=4cm,outer=3cm,top=2.5cm,bottom=3.5cm,pdftex]{geometry}

\usepackage{setspace} 

\usepackage{graphicx,xcolor}
\graphicspath{{images/}}

\usepackage{subfig}
\usepackage{booktabs}
\usepackage{lipsum}
\usepackage{microtype}
\usepackage{url}
\usepackage[final]{pdfpages}

\usepackage{fancyhdr}

\fancypagestyle{plain}{
	\fancyhf{}

	\fancyfoot[EL,OR]{\thepage}}
\fancypagestyle{addpagenumbersforpdfimports}{
	\fancyhead{}
	
	\fancyfoot{}
	\fancyfoot[RO,LE]{\thepage}
}

\usepackage{listings}
\usepackage{hyperref}
\usepackage[
section,
acronym,      
sort=def]       
{glossaries}

\newglossary[slg]{symbols}{syi}{syg}{Index of Symbols}

\makeglossaries

\hypersetup{pdfborder={0 0 0},
	colorlinks=false,
	linkcolor=black,
	citecolor=black,
	urlcolor=black}

\usepackage{makeidx}
\makeindex

\makeatletter
\def\cleardoublepage{\clearpage\if@twoside \ifodd\c@page\else
    \hbox{}
    \thispagestyle{empty}
    \newpage
    \if@twocolumn\hbox{}\newpage\fi\fi\fi}
\makeatother \clearpage{\pagestyle{plain}\cleardoublepage}

\usepackage{color}
\usepackage{tikz}
\usepackage[explicit]{titlesec}

\titlespacing*{\chapter}{0pt}{50pt}{30pt}
\titlespacing*{\section}{0pt}{13.2pt}{*0}  
\titlespacing*{\subsection}{0pt}{13.2pt}{*0}
\titlespacing*{\subsubsection}{0pt}{13.2pt}{*0}

\newcounter{myparts}
\newcommand*\partlabel{}
\titleformat{\part}[display]  
	{\normalfont\bfseries\Huge} 
	{\gdef\partlabel{\thepart\ }}     
 	{0pt} 
 	  {\setlength{\unitlength}{20mm}
	  \addtocounter{myparts}{1}
	  \begin{tikzpicture}[remember picture,overlay]
    \node[anchor=north west,xshift=-65mm,yshift=-6.9cm-\value{myparts}*20mm] at (current page.north east) 
      {\begin{tikzpicture}[remember picture, overlay]
        \draw[fill=black] (0,0) rectangle(62mm,20mm);   
        \node[anchor=north west,yshift=-6.1cm-\value{myparts}*20mm,xshift=-60.5mm,minimum height=30mm,inner sep=0mm] at (current page.north east)
        {\parbox[top][30mm][t]{55mm}{\raggedright \color{white}Part \partlabel $\phantom{\textrm{l}}$}};  
        \node[anchor=north east,yshift=-6.1cm-\value{myparts}*20mm,xshift=-63.5mm,text width=\textwidth,minimum height=30mm,inner sep=0mm] at (current page.north east)
              {\parbox[top][30mm][t]{\textwidth}{\raggedleft \color{black}#1}};
       \end{tikzpicture}
      };
   \end{tikzpicture}
   \gdef\partlabel{}
  } 

\usepackage{amsmath}
\makeatletter
\def\resetMathstrut@{%
  \setbox\z@\hbox{%
    \mathchardef\@tempa\mathcode`\(\relax
      \def\@tempb##1"##2##3{\the\textfont"##3\char"}%
      \expandafter\@tempb\meaning\@tempa \relax
  }%
  \ht\Mathstrutbox@1.2\ht\z@ \dp\Mathstrutbox@1.2\dp\z@
}
\makeatother

\usepackage{amssymb,eucal,graphicx,times,exscale,epsfig,amsfonts,amsthm,mathrsfs,fancybox,
latexsym,mathtools}
\usepackage{abc}
\usepackage{harmony}
\usepackage{wasysym}
\usepackage{calligra}
\usepackage[T1]{fontenc}
\usepackage{fixltx2e}
\usepackage{float}

\theoremstyle{remark}

\theoremstyle{definition}

\usepackage{titlepic}

\def\<{\langle}
\def\>{\rangle}

\usepackage{url}

\DeclarePairedDelimiterX{\infdivx}[2]{(}{)}{%
  #1\delimsize\| #2%
}

\usepackage{paralist}
        {\begin{itemize}[#1]}{\end{itemize}%
         \vspace{-.6\baselineskip}}

        {\vspace{-\baselineskip}\begin{list}{#1}{%
        \setlength{\partopsep}{0pt}%
        \setlength{\topsep}{0pt}}}
        {\end{list}\vspace{-.6\baselineskip}}

        {\begin{compactitem}[#1]}{\end{compactitem}}

\hypersetup{linktoc=none}

\begin{document}
\frontmatter
\begin{titlepage}
\begin{center}
\sffamily

\null\vspace{2cm}
{\Large Re-proving Channel Polarization Theorems:  \\[12pt] An Extremality and Robustness Analysis} \\[24pt] 
\textcolor{gray}{\small{THIS IS A TEMPORARY TITLE PAGE \\ It will be replaced for the final print by a version \\ provided by the service academique.}}
    
\vfill

	THÈSE N$^{\circ}$ 6403\\
	\vspace*{0.3cm}
	\small\textsc{présenté} \\
	\textsc{à la faculté informatique et communications}\\
	\textsc{laboratoire de théorie de l'information}\\
	\textsc{programme doctoral en informatique, communications et information}\\\normalsize
	\vspace*{0.5cm}
	ÉCOLE POLYTECHNIQUE FÉDÉRALE DE LAUSANNE\\
	\vspace*{0.5cm}
	\textsc{pour l'obtention du grade de docteur ès sciences}\\
	\vspace*{1cm}
	\small\textsc{par}\\ \normalsize
	
	\vspace*{0.5cm}
   \large Mine ALSAN\\  
\vspace*{2cm}
\small
acceptée le 18 Septembre 2014 sur proposition du 
jury:\\[4pt]
\vspace*{0.3cm}
    Prof Martin Odersky, président du jury\\
    Prof Emre Telatar, directeur de thèse\\
    Prof Erdal Ar{\i}kan, rapporteur\\
    Prof Albert Guill\'en i F\`abregas, rapporteur\\
    Prof Amos Lapidoth, rapporteur\\[12pt]

\vspace{0.5cm}
\includegraphics[width=4cm]{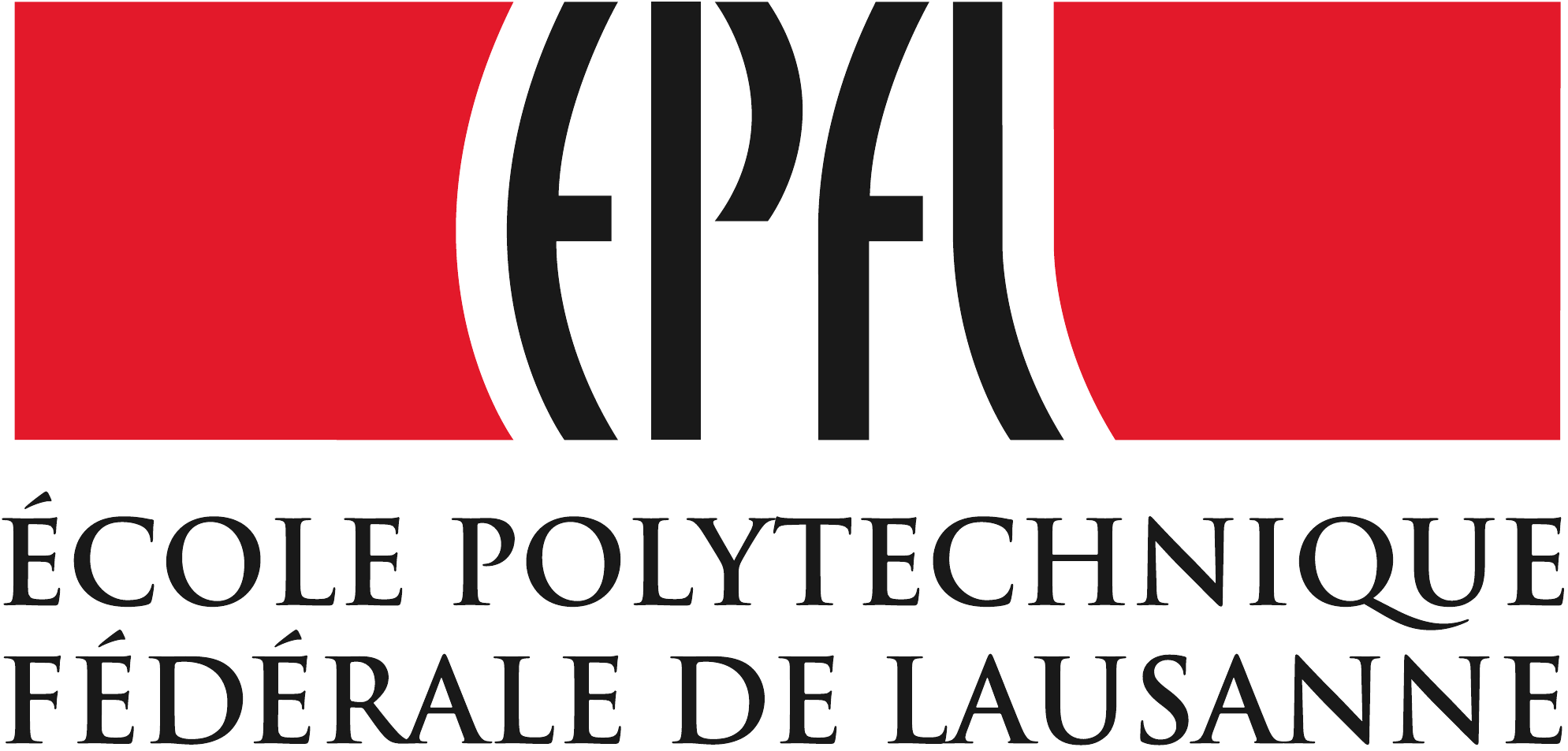}\\
Suisse, 2014\\
\end{center}

\end{titlepage}


\cleardoublepage
\pagenumbering{roman}
\setcounter{page}{11}
\cleardoublepage
\chapter*{Abstract}
\markboth{Abstract}{Abstract}

The general subject considered in this thesis is a recently discovered coding technique, polar coding, which is used to construct a class of error correction codes with unique properties. In his ground-breaking work, Ar{\i}kan proved that this class of codes, called polar codes, achieve the symmetric capacity --- the mutual information evaluated at the uniform input distribution --- 
of any stationary binary discrete memoryless channel with low complexity encoders and decoders 
requiring in the order of $O(N\log N)$ operations in the block-length $N$. 
This discovery settled the long standing open problem left by Shannon of finding low complexity codes achieving the channel capacity. 

Polar codes are not only appealing for being the first to `close the deal'. In contrast to most of the existing coding schemes, polar codes admit an explicit low complexity construction. In addition, for symmetric channels, the polar code construction is deterministic; the theoretically beautiful but practically limited ``average performance of an ensemble of codes is good, so there must exist one particular code in the ensemble at least as good as the average'' formalism of information theory is bypassed. Simulations are thus not necessary in principle for evaluating the error probability which is shown in a study by Telatar and Ar{\i}kan to scale exponentially in the square root of the block-length. As such, at the time of this writing, polar codes are appealing for being the only class of codes proved, and proved with mathematical elegance, to possess all of these properties.  

Polar coding settled an open problem in information theory, yet opened plenty of challenging problems that need to be addressed. 
This novel coding scheme is a promising method from which, in addition to data transmission, problems such as data compression or compressed sensing, 
which includes all types of measurement processes like the MRI or ultrasound, could benefit in terms of efficiency. To make this technique fulfil its promise, 
the original theory has been, and should still be, extended in multiple directions. A significant part of this thesis is dedicated to advancing the knowledge 
about this technique in two directions. The first one provides a better understanding of polar coding by generalizing some of the existing results 
and discussing their implications, and the second one studies the robustness of the theory over communication models introducing various forms of uncertainty or variations into the probabilistic model of the channel. 

The idea behind the design of a polar code is a phenomenon called channel polarization. 
This consists of synthesizing two new channels by applying the polar transform to two other channels. In the process, it is observed that while the sum symmetric capacities are preserved, 
the overall reliability is improved by creating `variance', i.e., the two new channels are created in such a way that the difference between their symmetric capacities is strictly larger than the difference between the symmetric capacities of the original pair of channels as long as the channels are not already perfect or completely noisy. Consequently, the new synthetic channels polarize: one becomes better and the other worse than the original mediocre channels. This result follows as a corollary to information combining which shows that the extremal bounds of the difference between the symmetric capacities of the created channels are attained by the binary erasure channel and the binary symmetric channel.

The mutual information, though fundamental, is not the only information measure of interest to the information theory community. In the field's literature, `Gallager's $E_0(\rho)$', for $\rho> -1$, is a well rooted family of information measures appearing in various error exponent problems and also in sequential decoding. The mutual information, determining the theoretical limit of information transmission, and the cutoff rate, another channel parameter which used to be interpreted as the `practical limit' of information transmission, turn out both to be special cases of $E_0(\rho)/\rho$. 
In retrospect, Ar{\i}kan's discovery came as the offspring of his prior work looking into a method to close the gap between the mentioned two limits. 

Based on this account, we study as part of this thesis the evolution of this more general family of information measures under the polar transform. In particular, we prove that the polar transform improves $E_0(\rho)$ for binary input channels. The result helps us understand better why the polar transform yields capacity achieving and low complexity codes: the improvement in $E_0(\rho)$ translates into an improvement in the complexity--error-probability trade-off. This is a concept introduced in the 1996 Shannon Lecture given by Forney. In addition, we prove that even if we change the measure of information from the customary mutual information to $E_0(\rho)$, the binary erasure channel and the binary symmetric channel still remain extremal. Speaking of extremality, we also show independent from any polarization context the extremality of these two channels amongst all binary input channels of a given $E_0(\rho)$ value evaluated at a fixed $\rho$. 

Once a deeper understanding of the technique of polar coding is developed, the thesis proceeds with the study of a practical problem related to the design of polar codes: ``robustness against channel parameter variations'', as stated in Ar{\i}kan's original work. Working out this problem is particularly challenging for polar coding as the initial development revealed that polar codes are channel specific designs. However, from an engineering point of view, it is critical that the results of a theory be robust. This is why right after its conception, partial orderings for channels became relevant for designing polar codes. Two channels are ordered if the code designed for one of the channels can be mapped to a code resulting in at most the same decoding error probability when used over the other channel. In fact, it was once more Shannon who introduced in a note the concept of partial orderings for discrete memoryless channels. In this thesis, we first touch this topic by introducing a rigorous framework in which we propose to study partial orderings for communication channels in the context of stochastic orders known as convex orderings. In this process, we discover a novel partial ordering for binary discrete memoryless channels we call the symmetric convex ordering. Then, the thesis focuses on different communication models proposed in the literature for building more robust systems; chapters are dedicated to extend the original theory of polar coding to the following complex scenarios:

Coding with a given decision rule--- In this scenario, we study the performance of mismatched polar decoders. A mismatched polar decoder is a polar successive cancellation decoder which uses, instead of the true channel's law, the metric of a mismatched channel during the decision procedure. We find the transmission capacity of polar coding with mismatched polar decoding. Moreover, we show that this capacity is lower bounded by a certain family of improving lower bounds converging to the polar mismatched capacity; whenever any of these bounds are positive, strictly positive communication rates can be achieved with properly constructed polar codes. We also observe that the block decoding error probability still decays exponentially in the square root of the block-length as in the matched case. It is worth emphasizing that while extending the theory of polar coding to mismatched communication scenarios, the mismatched polar decoder preserves the $O(N\log N)$ low complexity structure of the `matched' polar decoder. This structural advantage further motivates polar coding in the presence of a decoding mismatch.  

Communication over a class of channels--- We also investigate in this thesis the design of robust polar codes over a class of channels. Generally in this scenario, the code designer has access only to a partial knowledge about the true channel through the class to which it belongs. 
The problem is approached from different angles. First by allowing the decoder to know the true channel, we link polar ordering to the symmetric convex ordering, the novel order introduced by this thesis. Then letting instead the encoder know the channel, we extend the results about the mismatched capacity of polar codes to the compound setting by using the notion of one-sided sets of channels introduced by Abbe and Zheng. Taking yet another approach, we show that polar codes using an approximation at the decoder side are robust over the class of binary symmetric channels. Combining this result with simulations, we provide strong evidence that polar codes are `practically universal' over binary symmetric channels. Finally, we prove that universality can be traded for complexity by showing that multiple runs of the polar decoder implementing a generalized likelihood ratio test give a universal decoding rule for binary input channels satisfying certain mild conditions. Hence, more resources at the decoder is the price for universality. 

Communication over non-stationary channels--- A further original contribution of this thesis is the extension of the theory of channel polarization over non-stationary memoryless channels. This is a model which is quite useful to capture the effects of time-varying noise present in real communication systems as it is no longer assumed that the communication channel is stationary during the transmission of information. As the existing proof techniques are not applicable to this scenario, we first reprove the polarization phenomenon by using only elementary methods. Then by using the same method, we show that Arıkan’s construction also polarizes non-stationary memoryless channels in the same way it polarizes stationary ones.

\vskip0.5cm
\textbf{Key words:} Polar coding, polar codes, channel polarization, mismatched decoding, compound channels, robust code design, generalized likelihood ratio test (GRLT), coding for non-stationary channels, extremal channels, Gallager's $E_0$, error exponents, information combining.

\begin{otherlanguage}{french}
\cleardoublepage
\chapter*{Résumé}
\markboth{Résumé}{Résumé}
Le sujet principal de cette thèse est une technique de codage récemment découverte, le codage polaire, destinée à construire une famille de codes correcteurs aux propriétés uniques. Dans son travail de fondateur,  Ar{\i}kan a démontré que cette famille de codes correcteurs, appelés les codes polaires, atteignent la capacité symétrique---l'information mutuelle évaluée sous une distribution d'entrée uniforme--- de tout canal binaire discret sans mémoire et stationnaire avec des codeurs et des décodeurs à faible complexité exigeant de l'ordre de $O(N\log N)$ opérations en la longueur du bloc $N$. Cette découverte a résolu le problème laissé ouvert par Shannon d'inventer des codes qui atteignent la capacité avec une faible complexité.

Les codes polaires ne sont pas seulement intéressants parce qu'ils sont les premiers `à conclure l'affaire'. Contrairement à la plupart des systèmes de codage existants, les codes polaires admettent une construction explicite de faible complexité. De plus, pour les canaux symétriques, la construction de codes polaires est déterministe. La technique usuelle de la théorie de l'information attestant que ``si la performance moyenne d'un ensemble de codes est bonne, alors il doit y avoir au moins un code de l'ensemble aussi bon que la moyenne'' est belle en théorie mais limitée en pratique; les codes polaries contournent cette approche traditionnelle et des simulations ne sont pas en principe nécessaires pour évaluer la probabilité d'erreur d'un code donné. Une étude réalisée par Telatar et Ar{\i}kan indique que celle-ci décroît proportionnellement à l'exponentielle de la racine carrée de la longueur du bloc. A ce titre, au moment d'écrire ces lignes, les codes polaires sont la seule famille de codes démontrée, et démontrée avec élégance mathématique, à posséder toutes ces propriétés.

Le codage polaire a résolu un problème ouvert depuis longtemps en théorie de l'information mais, en même temps, a posé plusieurs problèmes difficiles qui doivent être abordés. Ce nouveau schéma de codage est une méthode prometteuse grâce à laquelle, en plus de la transmission de données, des problèmes tels que la compression de données ou l'acquisition comprimée, comprenant tous les types de processus de mesure comme l'IRM ou l'échographie, pourraient en bénéficier en efficacité. Afin de permettre à cette technique de tenir sa promesse, la théorie originale a été, et devra encore être, étendue dans plusieurs directions. Une partie considérable de cette thèse est consacrée à l'avancement des connaissances sur cette technique dans deux directions. La première permet une meilleure compréhension du codage polaire en généralisant certains résultats existants et en discutant leurs implications, et la seconde étudie la robustesse de la théorie par rapport aux modèles de communication introduisant diverses formes d'incertitude ou de variations dans le modèle probabiliste du canal.

L'idée derrière la conception de codes polaires est un phénomène appelé la polarisation de canal. Ce phénomène consiste en ceci: à synthétiser deux nouveaux canaux en appliquant la transformée polaire à deux autres canaux. Il est observé que, dans le procédé, la somme des capacités symétriques est conservée tandis que la fiabilité globale est améliorée par la création de `variance ', c'est-à-dire que les deux nouveaux canaux sont créés de manière à ce que la différence entre leurs capacités symétriques soit strictement plus grande que la différence entre les capacités symétriques des canaux originaux tant que ces derniers ne sont pas déjà  sans bruit ou complètement bruités. Par conséquent, les nouveaux canaux synthétiques sont polarisés: l'un devient meilleur et l'autre plus mauvais que les canaux médiocres du début. Ce résultat est un corollaire à `information combining' qui montre que les limites extrémals de la différence entre les capacités symétriques des nouveaux canaux sont atteintes par le canal binaire à effacement et le canal binaire symétrique.

L'information mutuelle, bien que fondamentale, n'est pas la seule mesure d'information d'intérêt pour la communauté de la théorie de l'information. Dans la littérature, les fonctions $E_0(\rho)$ introduites par Gallager, pour $\rho> -1$, constituent une famille de mesures d'information bien enracinée qui apparaît dans divers problèmes d'exposants d'erreurs et aussi dans le décodage séquentiel. L'information mutuelle, qui détermine la limite théorique de la transmission de l'information, et le taux de coupure, un autre paramètre de canal qui a été interprété autrefois comme la `limite pratique' de la transmission de l'information, se révèlent en tant que cas spéciaux de $E_0(\rho)/\rho$. En rétrospective, la découverte d'Ar{\i}kan est le résultat de sa recherche d'une méthode pour combler l'écart entre les deux limites mentionnées.

A cet égard, nous étudions dans le cadre de cette thèse l'évolution de cette famille de mesures d'information sous la transformée polaire. En particulier, nous démontrons que la transformée polaire améliore le paramètre $E_0(\rho)$ des canaux à entrées binaires. Le résultat nous permet de mieux comprendre la raison pour laquelle la transformée polaire donne des codes qui atteignent la capacité et qui sont de faible complexité: l'amélioration du paramètre  $E_0(\rho)$ se traduit par une amélioration du compromis entre la complexité et la probabilité d'erreur. Il s'agit d'un concept introduit en 1996 par Forney. De plus, nous démontrons que même si la mesure d'information est changée de l'information mutuelle habituelle à $E_0(\rho)$, le canal binaire à effacement et le canal binaire symétrique restent toujours extrémaux. En parlant des canaux extrémaux, nous caractérisons aussi, indépendamment de tout contexte de polarisation, l’extrémalité de ces deux canaux parmi tous les canaux à entrées binaires ayant une valeur donnée de $E_0(\rho)$ pour une valeur fixe de $\rho$.

Après avoir développé une compréhension plus profonde de la technique de codage polaire, la thèse procède à l'étude d'un problème pratique relié à la conception de codes polaires indiqué dans l'étude originale d'Ar{\i}kan: ``la robustesse contre les variations des paramètres du canal''. La résolution de ce problème est particulièrement difficile pour le codage polaire, parce que le développement initial a montré que la conception d'un code polaire est adaptée spécifiquement à la loi de distribution du canal de communication. Cependant, du point de vue de l’ingénieur, il est essentiel que les résultats d'une théorie soit robustes. Ainsi, juste après sa conception, les ordres partiels pour les canaux
sont devenus pertinents pour la conception de codes polaires.  
Deux canaux sont ordonnés si le code destiné à l'un des canaux peut être transformé en un code qui donne au maximum la même probabilité d'erreur de décodage si utilisé sur l'autre canal. Une fois de plus, c'est Shannon qui a introduit dans une note le concept d'ordres partiels pour les canaux discrets sans mémoire. Dans cette thèse, nous traitons ce sujet en introduisant un cadre rigoureux dans lequel nous proposons d'étudier les ordres partiels pour les canaux de communication dans le contexte d'ordres stochastiques appelés ordres convexes. Plus précisément, nous découvrons un nouvel ordre partiel pour les canaux binaires discrets sans mémoire que nous appelons l'ordre convexe symétrique. Par la suite, la thèse examine les différents modèles de communication proposés dans la littérature pour construire des systèmes plus robustes. Des chapitres de cette thèse sont dédiés à étendre la théorie originale du codage polaire aux scénarios complexes suivants:

Codage avec une règle de décision donnée--- Dans ce scénario, nous étudions la performance des décodeurs polaires désadaptés. Un décodeur polaire désadapté est un décodeur polaire à annulations successives qui utilise au cours de la procédure de décision la loi de distribution d'un canal désadapté au lieu de la loi du vrai canal. Nous définissons la capacité de transmission avec décodage polaire désadapté. De plus, nous montrons qu'il existe une famille de bornes inférieures à cette capacité et nous faisons la conjecture que la familles de bornes converge vers cette capacité lorsque la longueur du bloc devient grande. Donc, quand l'une de ces bornes est positive, des taux de communication strictement positifs peuvent être atteints avec des codes polaires appropriés. Nous observons également que la probabilité d'erreur de décodage du bloc décroît proportionnellement à l'exponentielle de la racine carrée de la longueur du bloc, comme précédemment. Il faut aussi souligner que, tout en étendant la théorie du codage polaire à des cas de communication avec des canaux désadaptés, le décodeur polaire désadapté préserve la même structure à faible complexité de l'ordre de $O(N\log N)$ que le décodeur polaire `adapté'. Cette structure à faible complexité motive davantage le codage polaire en présence de décodage désadapté.

Communication sur une famille de canaux--- Nous étudions également dans cette thèse la conception de codes polaires robustes sur une famille de canaux. Généralement, dans ce scénario, nous avons seulement accès à une connaissance partielle du vrai canal via la famille à laquelle il appartient. Donc, un code universel doit être conçu pour la famille de canaux. Le problème est abordé sous différents points de vue. D'abord, en permettant au décodeur (mais pas au codeur) de connaître le vrai canal, nous relions l'ordre polaire à l'ordre convexe symétrique, le nouvel ordre partiel introduit par cette thèse. Ensuite, en permettant au codeur, au lieu du décodeur, d'avoir connaissance du vrai canal, nous étendons les résultats de la thèse sur la capacité de transmission avec décodage polaire désadapté en utilisant la notion de famille de canaux unilatéraux introduite par Abbé et Zheng. Prenant encore une autre approche, nous montrons que les codes polaires utilisant une méthode de calcul approximative au décodeur sont robustes pour la famille de canaux binaires symétriques. En combinant ce résultat avec des simulations, nous fournissons des preuves solides qui montrent que les codes polaires sont `pratiquement universels' sur les canaux binaires symétriques. Enfin, nous démontrons que l'universalité peut être échangée contre la complexité. Nous montrons que plusieurs appels au décodeur polaire mettant en œuvre un test du rapport de vraisemblance généralisé donnent une règle de décodage universelle sur les canaux à entrées binaires qui satisfont certaines conditions. Par conséquent, il y a besoin de plus de ressources au niveau du décodeur pour atteindre l'universalité. 

Communication sur les canaux non-stationnaires--- Une autre contribution originale de cette thèse est l'extension de la théorie de la polarisation de canal aux canaux sans mémoire qui sont non-stationnaires. Ce modèle, qui ne suppose plus que le canal de communication est stationnaire durant la transmission de l'information, est très utile pour capter les effets des variations temporelles du bruit présent dans les systèmes de communication réels. Comme les techniques de preuve existantes ne sont pas applicables à ce scénario, nous reprouvons à nouveau le phénomène de polarisation pour le cas stationnaire en utilisant uniquement des méthodes élémentaires. Ensuite, en nous servant de la même méthode, nous montrons que la construction d'Arıkan polarise également les canaux non-stationnaires sans mémoire, de la même manière qu'elle polarise ceux qui sont stationnaires. 

\vskip0.5cm
\textbf{Mots clefs:}
Codage polaire, codes polaires, polarisation de canal,  décodage désadapté, famille de canaux, conception de code robuste, test du rapport de vraisemblance généralisé (GLRT), codage pour canaux non-stationnaires, canaux extrémaux, $E_0$ de Gallager, exposants d'erreur, `information combining'.
\end{otherlanguage}


\cleardoublepage
\chapter*{Contents}
\markboth{Contents}{}
\makeatletter
\input{my_thesis.toc}
\makeatother

%


\setlength{\parskip}{1em}

\mainmatter













\clearpage
\thispagestyle{empty}
\hspace{10mm}
\newpage
\pagenumbering{arabic}
\setcounter{page}{201}


\begin{thebibliography}{10}

\bibitem{6773024}
C.~E. Shannon.
\newblock A mathematical theory of communication.
\newblock {\em The Bell System Technical Journal}, 27(3):379--423, 1948.

\bibitem{1669570}
E.~Ar{\i}kan.
\newblock Channel polarization: A method for constructing capacity-achieving
  codes for symmetric binary-input memoryless channels.
\newblock {\em IEEE Trans. Inf. Theory}, 55(7):3051--3073, 2009.

\bibitem{6557004}
I.~Tal and A.~Vardy.
\newblock How to construct polar codes.
\newblock {\em IEEE Trans. Inf. Theory}, 59(10):6562--6582, 2013.

\bibitem{10.1109/TIT.2005.862081}
E.~Ar{\i}kan.
\newblock Channel combining and splitting for cutoff rate improvement.
\newblock {\em IEEE Trans. Inf. Theory}, 52(2):628--639, 2006.

\bibitem{578869}
R.~G. Gallager.
\newblock {\em Information Theory and Reliable Communication}.
\newblock John Wiley \& Sons, Inc., New York, NY, USA, 1968.

\bibitem{4282117}
D.~J. Costello and G.~D. Forney~Jr.
\newblock Channel coding: The road to channel capacity.
\newblock {\em Proc. of the IEEE}, 95(6):1150--1177, 2007.

\bibitem{1057827}
R.~Fano.
\newblock A heuristic discussion of probabilistic decoding.
\newblock {\em IEEE Trans. Inf. Theory}, 9(2):64--74, 1963.

\bibitem{Elias}
P.~Elias.
\newblock Coding for noisy channels.
\newblock {\em IRE Conv. Rec.}, pages 37--46, 1955.

\bibitem{481781}
E.~Ar{\i}kan.
\newblock An inequality on guessing and its application to sequential decoding.
\newblock {\em IEEE Trans. Inf. Theory}, 42(1):99--105, 1996.

\bibitem{article:compound}
D.~Blackwell, L.~Breiman, and A.~J. Thomasian.
\newblock The capacity of a class of channels.
\newblock {\em The Annals of Mathematical Statistics}, 3(4):1229--1241, 1959.

\bibitem{csiszar1981information}
I.~Csisz{\'a}r and J.~K{\"o}rner.
\newblock {\em Information Theory: Coding Theorems for Discrete Memoryless
  Systems}.
\newblock Academic Press, Inc., Orlando, FL, USA, 1982.

\bibitem{394641}
I.~Csisz{\'a}r and P.~Narayan.
\newblock Channel capacity for a given decoding metric.
\newblock {\em IEEE Trans. Inf. Theory}, 41(1):35--43, 1995.

\bibitem{720546}
I.~Csisz{\'a}r.
\newblock The method of types [information theory].
\newblock {\em IEEE Trans. Inf. Theory}, 44(6):2505--2523, 1998.

\bibitem{article02}
S.~Arimoto.
\newblock Information measures and capacity of order $\alpha$ for discrete
  memoryless channels.
\newblock In I.~Csisz\'{a}r and P.~Elias, editors, {\em Topics in information
  theory}, volume~16, pages 41--52, The Netherlands, 1977. North-Holland
  Publishing Co.

\bibitem{370121}
I.~Csisz{\'a}r.
\newblock Generalized cutoff rates and {R}enyi's information measures.
\newblock {\em IEEE Trans. Inf. Theory}, 41(1):26--34, 1995.

\bibitem{6016020}
A.~R\'{e}nyi.
\newblock On measures of entropy and information.
\newblock {\em Proc. Fourth Berkeley Symp. on Math. Statist. and Prob.},
  1:547--561, 1961.

\bibitem{1053730}
R.~G. Gallager.
\newblock A simple derivation of the coding theorem and some applications.
\newblock {\em IEEE Trans. Inf. Theory}, 11(1):3--18, 1965.

\bibitem{1055007}
S.~Arimoto.
\newblock On the converse to the coding theorem for discrete memoryless
  channels (corresp.).
\newblock {\em IEEE Trans. Inf. Theory}, 19(3):357--359, 1973.

\bibitem{Massey94guessingand}
J.~L. Massey.
\newblock Guessing and entropy.
\newblock In {\em Proc. of the IEEE Int. Symposium on Inf. Theory}, page 204,
  1994.

\bibitem{notes1}
E.~Ar{\i}kan and E.~Telatar.
\newblock {BEC} and {BSC} are ${E}_{0}$ extremal.
\newblock Unpublished note.

\bibitem{dictionary}
{\em The American Heritage Dictionary of the English Language,}.
\newblock Fifth Edition copyright 2014 by Houghton Mifflin Harcourt Publishing
  Company.

\bibitem{6034105}
A.~Guill\'en~i F\`abregas, I.~Land, and A.~Martinez.
\newblock Extremes of random coding error exponents.
\newblock In {\em Proc. of the IEEE Int. Symposium on Inf. Theory}, pages
  2896--2898, 2011.

\bibitem{6377299}
A.~Guill\'en~i F\`abregas, I.~Land, and A.~Martinez.
\newblock Extremes of error exponents.
\newblock {\em IEEE Trans. Inf. Theory}, 59(4):2201--2207, 2013.

\bibitem{arXiv:Rate-of-pol}
E.~Ar{\i}kan and E.~Telatar.
\newblock On the rate of channel polarization.
\newblock {\em eprint arXiv:0807.3806}, 2008.

\bibitem{1412027}
I.~Sutskover, S.~Shamai, and J.~Ziv.
\newblock Extremes of information combining.
\newblock {\em IEEE Trans. Inf. Theory}, 51(4):1313--1325, 2005.

\bibitem{910572}
F.~R. Kschischang, B.~J. Frey, and H.-A. Loeliger.
\newblock Factor graphs and the sum-product algorithm.
\newblock {\em IEEE Trans. Inf. Theory}, 47(2):498--519, 2001.

\bibitem{Martingales:Book}
D.~Williams.
\newblock {\em Probability with Martingales}.
\newblock Cambridge mathematical textbooks. Cambridge University Press, 1991.

\bibitem{Forney:Lecture95}
G.~D. Forney~Jr.
\newblock 1995 {S}hannon {L}ecture—{P}erformance and complexity.
\newblock {\em IEEE. Inf. Theory Soc. Newslett.}, 46:3--4, 1996.

\bibitem{Viterbi:PDC}
A.~J. Viterbi and J.~K. Omura.
\newblock {\em Principles of Digital Communication and Coding}.
\newblock McGraw-Hill, New York, NY, USA, 1979.

\bibitem{5205856}
E.~Ar{\i}kan and I.~E. Telatar.
\newblock On the rate of channel polarization.
\newblock In {\em Proc. of the IEEE Int. Symposium on Inf. Theory,}, pages
  1493--1495, 2009.

\bibitem{5166430}
R.~Mori and T.~Tanaka.
\newblock Performance of polar codes with the construction using density
  evolution.
\newblock {\em IEEE Communications Letters}, 13(7):519--521, July 2009.

\bibitem{stanford1962generalized}
S.~Karlin and A.~Novikoff.
\newblock {\em Generalized Convex Inequalities}.
\newblock Pacific J. Math, 1963.

\bibitem{szekli1995stochastic}
R.~Szekli.
\newblock {\em Stochastic ordering and dependence in applied probability}.
\newblock Lecture notes in statistics. Springer-Verlag, 1995.

\bibitem{4461/THESES}
S.~B. Korada.
\newblock {\em Polar codes for channel and source coding}.
\newblock PhD thesis, Lausanne, 2009.

\bibitem{Blackwell1953}
D.~Blackwell.
\newblock Equivalent comparisons of experiments.
\newblock {\em The Annals of Mathematical Statistics}, 24(2):265--272, 1953.

\bibitem{Hurlimann2008}
W.~Hürlimann.
\newblock Extremal moment methods and stochastic orders.
\newblock {\em Boletín de la Asociación Matemática Venezolana},
  15(2):153--301, 2008.

\bibitem{6033724}
R.~Pedarsani, S.~H. Hassani, I.~Tal, and I.~E. Telatar.
\newblock On the construction of polar codes.
\newblock In {\em Proc. of the IEEE Int. Symposium on Inf. Theory}, pages
  11--15, 2011.

\bibitem{MMI:Goppa}
V.~D. Goppa.
\newblock Nonprobabilistic mutual information without memory.
\newblock {\em Probl. Contr. Inf. Theory}, 4:97--102, 1975.

\bibitem{1056798}
C.~E. Shannon.
\newblock The zero error capacity of a noisy channel.
\newblock {\em IRE Trans. on Inf. Theory}, 2(3):8--19, 1956.

\bibitem{Balakirsky}
V.~B. Balakirsky.
\newblock Coding theorem for discrete memoryless channels with given decision
  rule.
\newblock {\em Proc. of the First French-Soviet Workshop on Algebraic Coding},
  pages 142--150, Jul. 1991.

\bibitem{476314}
V.~B. Balakirsky.
\newblock A converse coding theorem for mismatched decoding at the output of
  binary-input memoryless channels.
\newblock {\em IEEE Trans. Inf. Theory}, 41(6):1889--1902, 1995.

\bibitem{Fischer}
T.~R.~M. Fischer.
\newblock Some remarks on the role of inaccuracy in {S}hannon's theory of
  information transmission.
\newblock In {\em Trans. of the Eighth Prague Conference}, volume~8A of {\em
  Czechoslovak Academy of Sciences}, pages 211--226. Springer Netherlands,
  1978.

\bibitem{kaplan1993information}
G.~Kaplan and S.~Shamai.
\newblock Information rates and error exponents of compound channels with
  application to antipodal signaling in a fading environment.
\newblock {\em A\"{E}U}, 47(4):228--239, 1993.

\bibitem{340469}
N.~Merhav, G.~Kaplan, A.~Lapidoth, and S.~Shamai.
\newblock On information rates for mismatched decoders.
\newblock {\em IEEE Trans. Inf. Theory}, 40(6):1953--1967, 1994.

\bibitem{abbeonesided}
E.~Abbe and L.~Zheng.
\newblock Linear universal decoding for compound channels.
\newblock {\em IEEE Trans. Inf. Theory}, 56(12):5999--6013, 2010.

\bibitem{6875075}
D.~Sutter and J.~M. Renes.
\newblock Universal polar codes for more capable and less noisy channels and
  sources.
\newblock In {\em Proc. of the IEEE Int. Symposium on Inf. Theory}, pages
  1461--1465, 2014.

\bibitem{less_noisy}
J.~K{\"o}rner and K.~Marton.
\newblock A source network problem involving the comparison of two channels.
\newblock {\em Trans. Colloq. Inf. Theory}, 1975.

\bibitem{5946819}
C.~Leroux, I.~Tal, A.~Vardy, and W.~J. Gross.
\newblock Hardware architectures for successive cancellation decoding of polar
  codes.
\newblock In {\em IEEE Int. Conference on Acoustics, Speech and Signal
  Processing}, pages 1665--1668, 2011.

\bibitem{CIT-004}
I.~Csiszár and P.~C. Shields.
\newblock Information theory and statistics: A tutorial.
\newblock {\em Foundations and Trends in Communications and Information
  Theory}, 1(4):417--528, 2004.

\bibitem{elgamal-kim:11}
A.~E. Gamal and Y.~H. Kim.
\newblock {\em Network Information Theory}.
\newblock Cambridge University Press, New York, NY, USA, 2011.

\bibitem{bollobas:86}
B.~Bollobás.
\newblock {\em Combinatorics}.
\newblock Cambridge University Press, Cambridge, 1986.

\bibitem{Wyner:Ziv}
A.~D. Wyner and J.~Ziv.
\newblock A theorem on the entropy of certain binary sequences and
  applications--i.
\newblock {\em IEEE Trans. Inf. Theory}, 19(6):769--772, 1973.

\bibitem{Info:combining}
I.~Land, S.~Huettinger, P.~A. Hoeher, and J.~B. Huber.
\newblock Bounds on information combining.
\newblock {\em IEEE Trans. Inf. Theory}, 51(2):612--619, 2005.

\bibitem{6033904}
I.~Tal and A.~Vardy.
\newblock List decoding of polar codes.
\newblock In {\em Proc. of the IEEE Int. Symposium on Inf. Theory}, pages 1--5,
  2011.

\bibitem{Shannon1958390}
C.~E. Shannon.
\newblock A note on a partial ordering for communication channels.
\newblock {\em Information and Control}, 1(4):390 -- 397, 1958.

\bibitem{privateEmreTelatar}
E.~Telatar.
\newblock Private communications.

\end{thebibliography}
\end{document}